\documentclass[9pt,conference]{IEEEtran}
\usepackage{amssymb,amsthm,amsmath,array}
\usepackage{graphicx}
\usepackage[caption=false,font=footnotesize]{subfig}
\usepackage{xspace}
\usepackage[sort&compress, numbers]{natbib}
\usepackage{stmaryrd}
\usepackage{xcolor}
\usepackage{mathtools}
\usepackage{float}
\usepackage{textcomp}

\DeclareRobustCommand{\IEEEauthorrefmark}[1]{\smash{\textsuperscript{\footnotesize #1}}}

\begin{document}
\title{Compressive Image Scanning Microscope}
\author{\IEEEauthorblockN{
        Ajay Gunalan\IEEEauthorrefmark{1,2},
        Marco Castello\IEEEauthorrefmark{3},
        Simonluca Piazza\IEEEauthorrefmark{3},
        Shunlei Li\IEEEauthorrefmark{1,2},
        Alberto Diaspro\IEEEauthorrefmark{4,5},
        Leonardo S. Mattos\IEEEauthorrefmark{1},
        Paolo Bianchini\IEEEauthorrefmark{4,5}
    }
    \IEEEauthorblockA{
        \IEEEauthorrefmark{1}Department of Advanced Robotics, Istituto Italiano di Tecnologia, Genoa, Italy\\
        \IEEEauthorrefmark{2}Department of Informatics, Bioengineering, Robotics and Systems Engineering, University of Genoa, Genoa, Italy \\
        \IEEEauthorrefmark{3}Genoa Instruments s.r.l., Genoa, Italy\\
        \IEEEauthorrefmark{4}Nanoscopy, Istituto Italiano di Tecnologia, Genoa, Italy\\
        \IEEEauthorrefmark{5}DIFILAB, Department of Physics, University of Genoa, Genoa, Italy
        }
}
\maketitle
\begin{abstract}
we present a novel approach to implement compressive sensing in laser scanning microscopes (LSM), specifically in image scanning microscopy (ISM), using a single-photon avalanche diode (SPAD) array detector. Our method addresses two significant limitations in applying compressive sensing to LSM: the time to compute the sampling matrix and the quality of reconstructed images. We employ a fixed sampling strategy, skipping alternate rows and columns during data acquisition, which reduces the number of points scanned by a factor of four and eliminates the need to compute different sampling matrices. By exploiting the parallel images generated by the SPAD array, we improve the quality of the reconstructed compressive-ISM images compared to standard compressive confocal LSM images. Our results demonstrate the effectiveness of our approach in producing higher-quality images with reduced data acquisition time and potential benefits in reducing photobleaching.
\end{abstract}

\section{Introduction}
Compressive sensing allows the reconstruction of high-dimensional ($N$) signals $x$ from low-dimensional ($M$) measurements $y$, as long as the signal is sparse on a particular basis, such as wavelet or shearlet ($N > M$). Generally, solving for $x$ in $y=Ax$ is an ill-posed problem, meaning there is no unique solution or the solution is not robust to small data perturbations \cite{EstrelaTotal}. Therefore, a regularizer term ($\phi$) is introduced to solve it.

\begin{equation}
\label{eq:cs}
\arg \min_{x} \;\; \phi(x) \;\;\; s.t. \;\;\;  Ax=y
\end{equation}

\begin{figure}[htbp]
\centerline{\includegraphics[width=\linewidth/\real{1.5}]{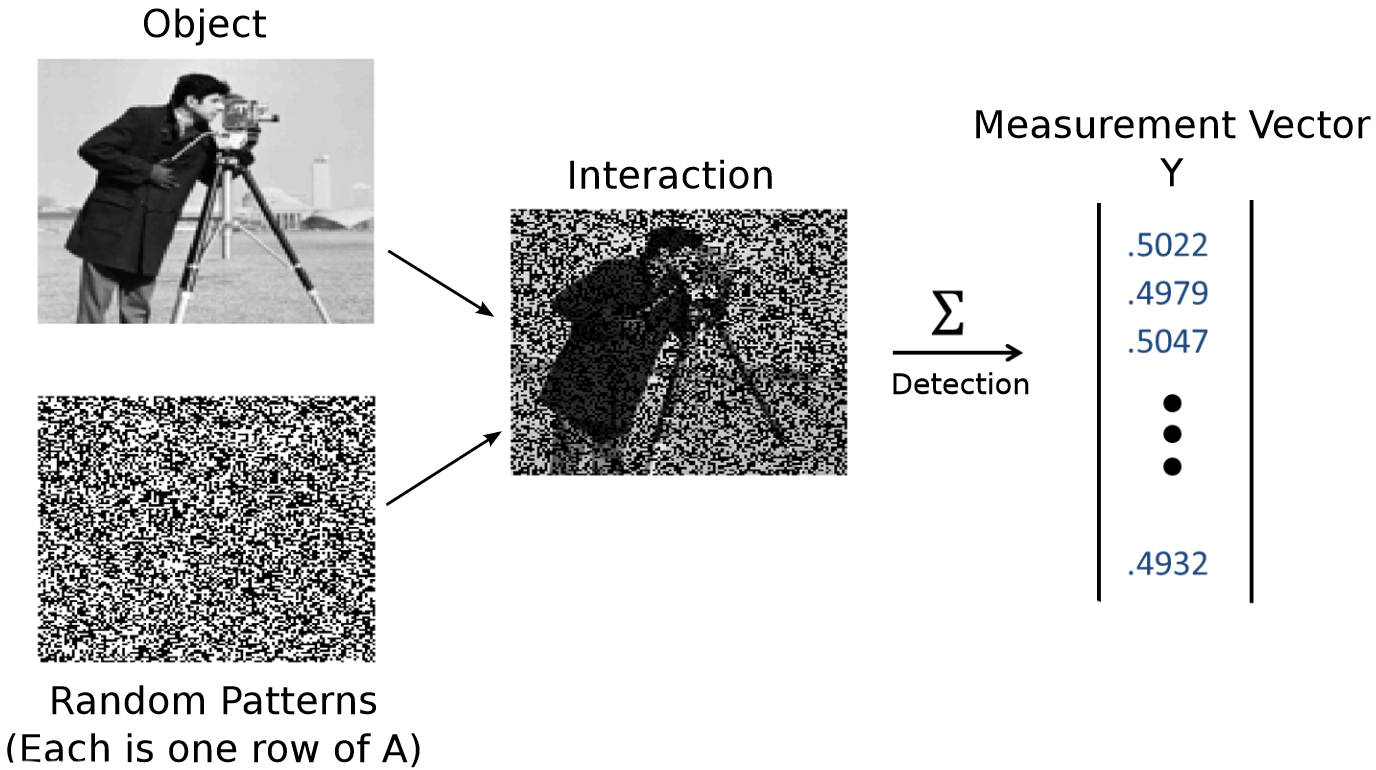}}
\caption{Standard optical architecture for compressive sensing based on the single-pixel camera.}
\label{stdCS}
\end{figure}

Compressive sensing is commonly implemented using a Digital Micromirror Device (DMD) or coded aperture in single-pixel cameras. Each row of $A$ corresponds to a unique binary mask, as illustrated in Fig.~\ref{stdCS}. $M$ unique binary masks sample the object of interest sequentially to obtain each element of the measurement matrix $Y$ \cite{Marcia2011Compressed}, \cite{Duarte2008Singlepixel}.

In Laser Scanning Microscopes (LSM), images are formed by scanning point-by-point, which is the basis for multiple imaging modalities like confocal microscopy, image scanning microscopy (ISM), and optical coherence tomography. Compressive sensing could enable higher temporal resolution and reduced photobleaching through efficient sampling. However, most existing laser scanning hardware doesn't use encoding devices like DMD or coded aperture, limiting the application of compressive sensing in such configurations.

Pavilion (2016) was the first to implement compressive sensing on such a configuration \cite{Pavillon2016Compressed}, using the point spread function (PSF) of the optical setup as a smoothing function, which led to a reduction of confocal fluorescence measurements by 10-15 times. In 2018, Francis \textit{et al.} \cite{Francis2018Multiresolutionbased} used a multi-resolution approach for better reconstruction in confocal images, but the quality, and time, of the reconstructed image is lower than std. TVAL3 solver. In 2021, Hu \textit{et al.} improved the speed in Raman imaging by sampling only the region of interest and avoiding scanning the background substrate, but the sampling matrix $A$ needed to be recomputed for every new image \cite{Hu2021Fast}. The practical application of compressive sensing to laser scanning imaging modalities is limited by three main factors: (1) the solver's reconstruction time, (2) the time to compute the sampling matrix ($A$), and (3) the quality of the reconstructed images.

In this work, we address the latter two issues. First, we skip alternate rows and columns during data acquisition, reducing the number of scanned points by a quarter, as illustrated in Fig.~\ref{sampling}. We use a fixed sampling matrix $A$ for different images, eliminating the need to compute different $A$ matrices. Secondly, we exploit the parallel images generated by the Image Scanning Microscope (ISM) on the single photon avalanche diode (SPAD) array detector \cite{spad} to improve the quality of reconstructed images. To our knowledge, this is the first work to implement compressive sensing on an image-scanning microscope (ISM).

\section{Materials and Methods}

\subsection{Simulation Setup and Ground Truth}
\begin{figure}[htbp]
\centerline{\includegraphics[width=\linewidth/\real{1}]{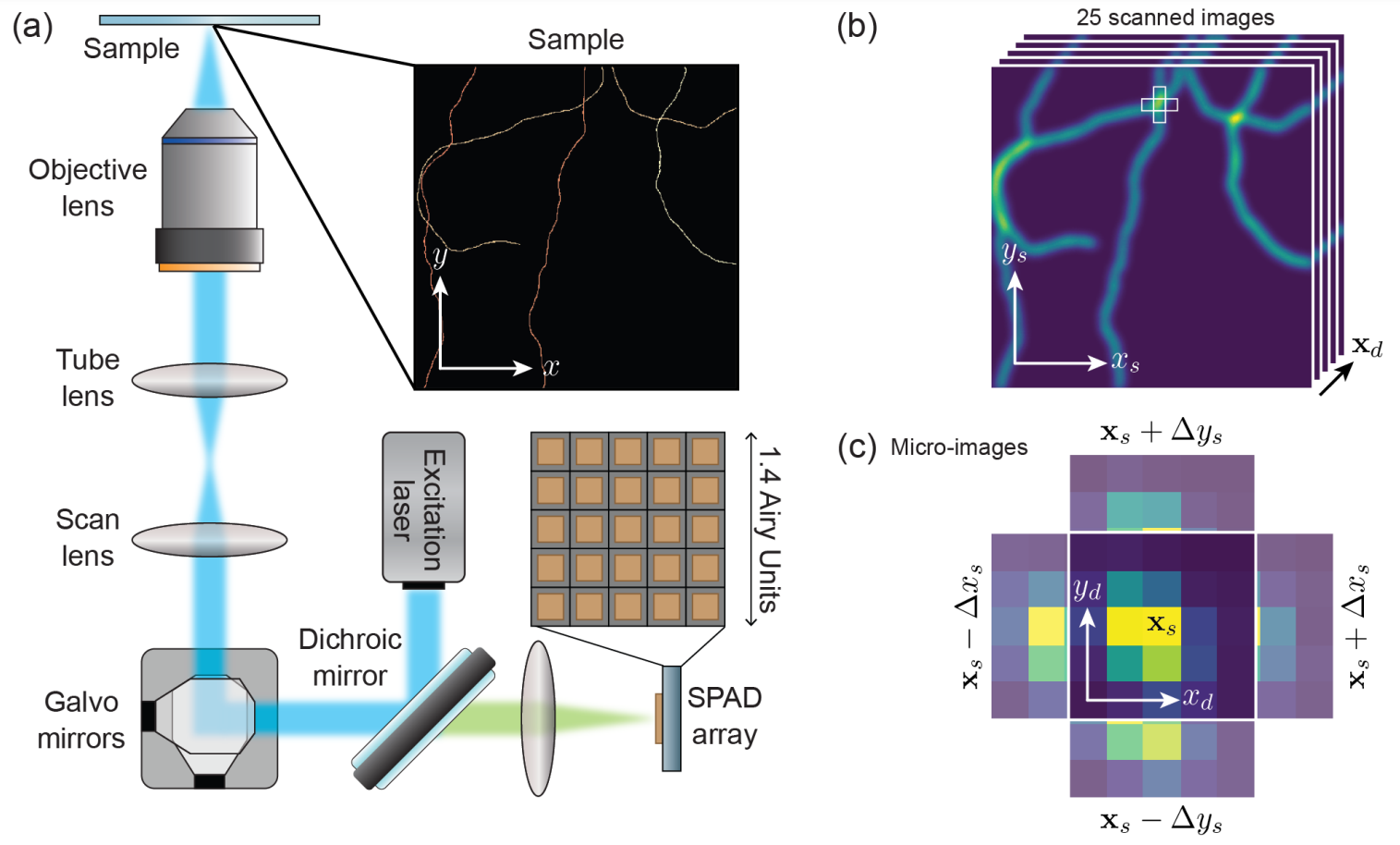}}
	\caption{Image Scanning Microscopy. (a) A sketch of the laser scanning microscope
equipped with a SPAD array detector. (b) The ISM dataset, seen as a set of scanned
images as many as the number of elements of the detector array. (c) The ISM dataset,
seen as a collection of micro-images, as many as the scan points. The depicted microimages correspond to the scan points highlighted in (b) as white boxes.}
	\label{ISM}
\end{figure}

We utilize an open-source ISM simulation software called BrightEyes-ISM to create ISM images. First, we generate a 2D point spread function (PSF) for each element in the SPAD array detector. The simulation space's pixel size is set to 25 nm, the detector element size to 50 nm, the detector element pitch to 75 nm, and the total magnification of the optical system to 500. The PSF is simulated for excitation and emission wavelengths of 640 nm and 660 nm, respectively. Next, we convolve the tubulin phantoms with the PSF for each element of the SPAD array detector to generate several parallel images, as shown in Fig.~\ref{ISM}, and add Poisson noise to the resulting images.

\subsection{Compressive Sensing Reconstruction}
There are two common choices for the regularizer $\phi(x)$ in Eq. \ref{eq:cs} \cite{Farnell2019Total}: (1) $L_1$ norm $||x||_{1}$ and (2) Total Variation ($TV$) norm $||x||_{TV}$:

\begin{equation}
||x||_{TV} :=\sum_{i=1}^{n_1}\sum_{j=1}^{n_2} |x_{i+1,j}-x_{i,j}|+ |x_{i,j+1}-x_{i,j}|
\end{equation}

TV regularization is more appropriate for image reconstruction because it preserves edges and boundaries \cite{ZhangEfficienta}. A comprehensive review of various algorithms for solving $L_1$ norm and $TV$ norm can be found in \cite{Sher2019Review}. Based on this information, we chose TVAL3 \cite{Li2013} due to its fast reconstruction time.

\begin{figure}[htbp]
\centerline{\includegraphics[width=\linewidth/\real{5}]{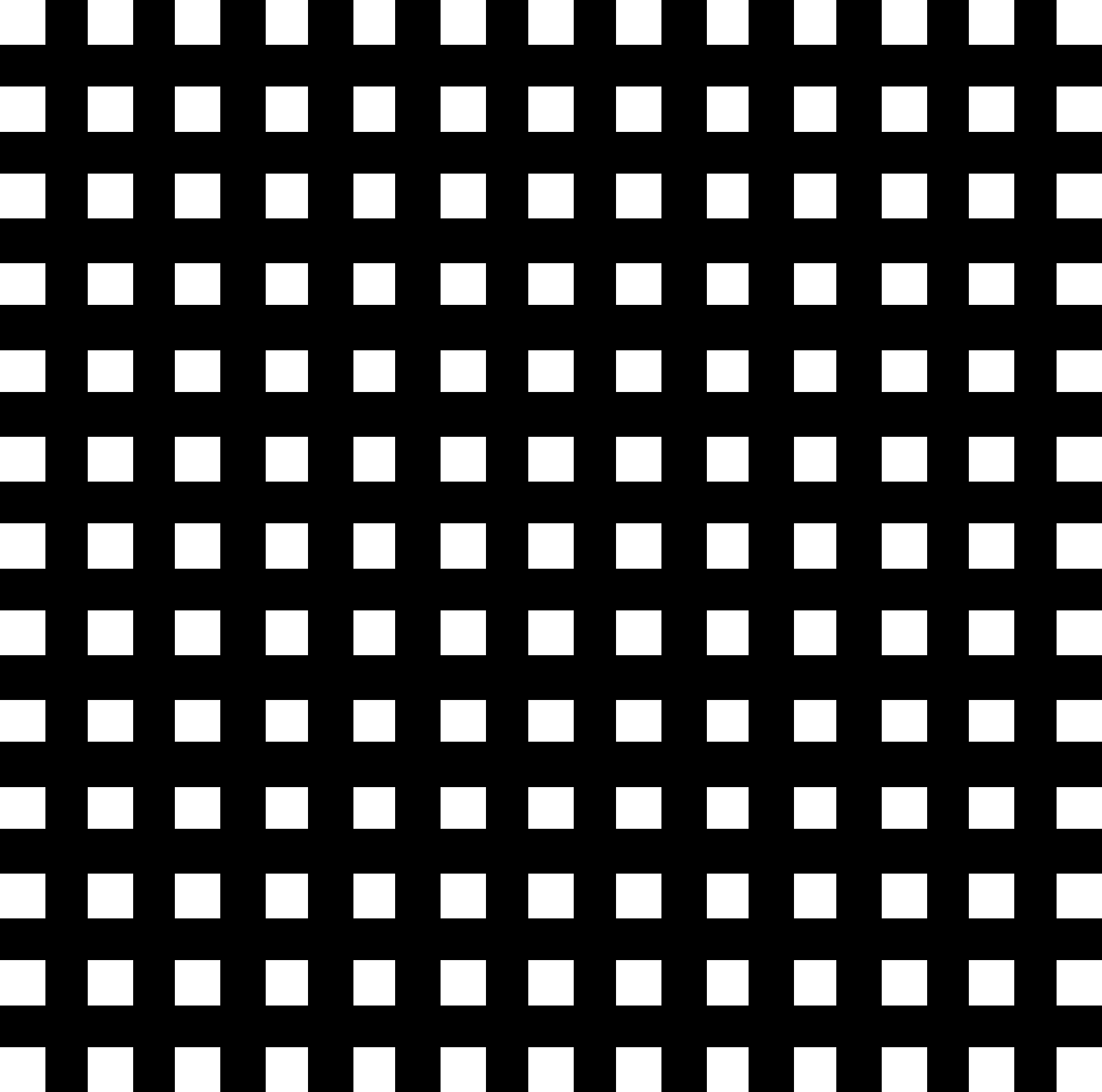}}
	\caption{Sampling strategy: Alternate rows and columns are skipped. White indicates sampled location, and black shows the unsampled location.}
	\label{sampling}
\end{figure}

The sampling matrix is designed to skip alternate rows and columns, as illustrated in Fig.~\ref{sampling}. This sampling matrix remains constant when the sample changes, so it only needs to be computed once. Following this, the sampled data ($y$) and sampling matrix ($A$) are employed to reconstruct the image using the TVAL3 solver.

\subsection{ISM Reconstruction}
The ISM image is generated by combining all parallel images from the SPAD array detector. We employ Adaptive Pixel Reassignment (APR) \cite{Castello2015}, \cite{Castello2019}, available in the BrightEyes-ISM package, for this merging process. The same procedure is applied to the corresponding reconstructed images from compressive sensing to obtain the Compressive-ISM images. For the Confocal LSM image, the central element of the SPAD array corresponds to the standard image, and in this case, the reconstructed image from compressive sensing is used directly.

\section{Results and Discussion}
\begin{figure}[htbp]
\centerline{\includegraphics[width=\linewidth/\real{1.5}]{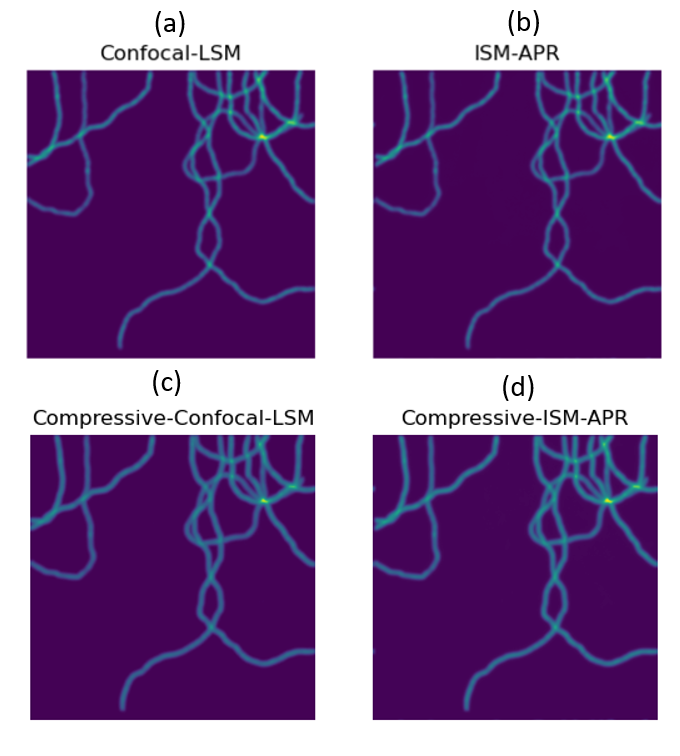}}
\caption{(a) Fully sampled CLSM. (b) Fully sampled ISM. (c) Compressive CLSM. (d) Compressive ISM.}
\label{reconstruction}
\end{figure}

We evaluate the quality of reconstructed images for both Compressive ISM and Compressive Confocal LSM by computing the relative error using Eq. \ref{eq:error}:

\begin{equation}
\text{Relative Error} = \frac{\| I_{fs} - I_{cs} \|_\text{F}}{\| I_{fs} \|_\text{F}} 
\label{eq:error}
\end{equation}

where $\| \|_\text{F}$  denotes the Frobenius norm, $I_{fs}$ represents the fully sampled image corresponding to the top row in Fig.\ref{reconstruction}, and $I_{cs}$ corresponds to the compressive images corresponding to the bottom row of Fig.\ref{reconstruction}. We used 25 samples to calculate the mean and standard deviation, as shown in Table \ref{tab:results}. Compressive ISM yields better results compared to Compressive Confocal LSM images, as illustrated in Fig.\ref{reconstruction}. This improvement can be attributed to the utilization of parallel images generated by the SPAD array. Moreover, our proposed sampling strategy reduces the number of scanned points by a factor of four, leading to faster data acquisition, decreased photobleaching in the sample, and eliminating the need to compute the sampling matrix $A$ for different samples.


\begin{table}[htbp]
\caption{Quality of reconstructed Images}
\begin{center}
\begin{tabular}{|c|c|c|c|}
 \hline
\textbf{Imaging Technique} & \textbf{Relative Error} \\ 
 \hline
Compressive Confocal LSM & $14.65\pm2.67 \; \%$ \\
 \hline
Compressive ISM  & $12.50\pm0.71 \; \%$ \\
 \hline
\end{tabular}
\label{tab:results}
\end{center}
\end{table}

\section{Conclusion}
We devised an effective compressive sensing method for image scanning microscopy using a SPAD array, employing a fixed sampling strategy to reduce acquisition time and negate the need for various sampling matrices. Our Compressive-ISM reconstruction provides better quality images than standard Compressive Confocal LSM. This work paves the way for practical applications in laser scanning imaging modalities, improving temporal resolution, minimizing photobleaching, and enhancing image quality. Future research includes hardware implementation, block-based compressive sensing \cite{Zammit2020} for faster reconstruction, GPU-based solutions for parallel SPAD array reconstruction, and deconvolution-based ISM algorithms \cite{https://doi.org/10.48550/arxiv.2211.12510} to further improve image quality.

\bibliographystyle{IEEEtran}
\bibliography{references}
\end{document}